\documentclass[letterpaper,english,aps, prl, reprint, superscriptaddress]{revtex4-2}
\usepackage[T1]{fontenc}
\usepackage[latin9]{inputenc}
\usepackage{verbatim}
\usepackage{amsmath}
\usepackage{amssymb}
\usepackage{graphicx}
\usepackage{relsize}
\usepackage{esint}
\usepackage{multirow}
\usepackage{adjustbox}
\usepackage{mathrsfs}

\usepackage{hyperref}
\hypersetup{
	colorlinks = true,
	allcolors = blue
}

\usepackage{times}

\makeatletter

\pdfpageheight\paperheight
\pdfpagewidth\paperwidth

\PassOptionsToPackage{position=top}{subfig}

\makeatother

\let\vec\boldsymbol
\usepackage{babel}
\usepackage{mathrsfs}

\draft

\begin{document}
\title{Fractional quantum anomalous Hall phase for Raman superarray of Rydberg atoms}

\author{Ting Fung Jeffrey Poon}
\thanks{The authors make equal contributions.}
\affiliation{International Center for Quantum Materials, School of Physics, Peking University, Beijing 100871, China}
\affiliation{Hefei National Laboratory, Hefei 230088, China}

\author{Xin-Chi Zhou}
\thanks{The authors make equal contributions.}
\affiliation{International Center for Quantum Materials, School of Physics, Peking University, Beijing 100871, China}
\affiliation{Hefei National Laboratory, Hefei 230088, China}

\author{Bao-Zong Wang}
\affiliation{International Center for Quantum Materials, School of Physics, Peking University, Beijing 100871, China}
\affiliation{Hefei National Laboratory, Hefei 230088, China}

\author{Tian-Hua Yang}
\affiliation{International Center for Quantum Materials, School of Physics, Peking University, Beijing 100871, China}
\affiliation{Hefei National Laboratory, Hefei 230088, China}

\author{Xiong-Jun Liu}
\email{Corresponding author: xiongjunliu@pku.edu.cn}
\affiliation{International Center for Quantum Materials, School of Physics, Peking University, Beijing 100871, China}
\affiliation{Hefei National Laboratory, Hefei 230088, China}
\affiliation{International Quantum Academy, Shenzhen 518048, China}
\affiliation{CAS Center for Excellence in Topological Quantum Computation, University of Chinese Academy of Sciences, Beijing 100190, China}

\begin{abstract}
Rydberg atom arrays offer promising platforms for quantum simulation of correlated quantum matter and raise great interests. This work proposes a novel stripe-lattice model with {\em Raman superarray of Rydberg atoms} to realize bosonic fractional quantum anomalous Hall (FQAH) phase. Two types of Rydberg states, arranged in a supperarray configuration and with Raman-assisted dipole-exchange couplings, are implemented to realize a minimal QAH model for hard-core bosons populated into a  topological flat band with large bulk gap under proper tunable experimental condition. With this the bosonic FQAH phase can be further achieved and probed feasibly. In particular, a novel quench protocol is proposed to probe the fractionalized excitations by measuring the correlated quench dynamics featured by fractional charge tunneling between bulk and chiral edge modes in the open boundary.
\end{abstract}

\maketitle
\begin{figure*}[tp]
\begin{center}
\includegraphics{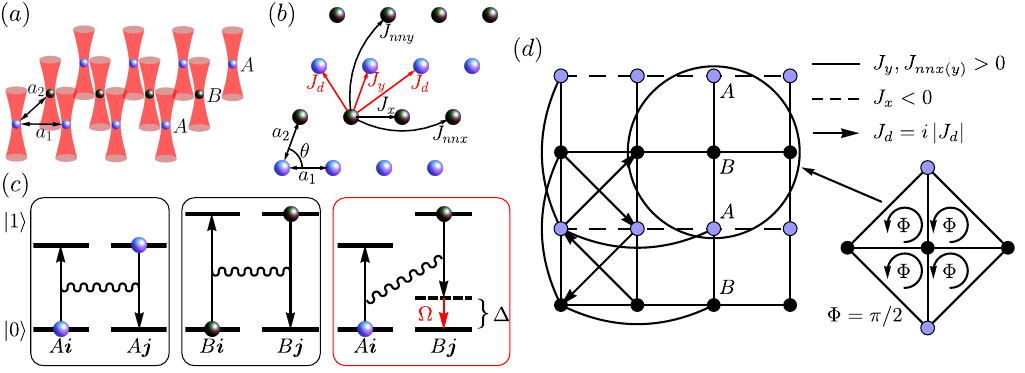}
\end{center}
\caption{\label{fig:SchemeFig}Sketch of the proposal and the stripe-lattice model. (a) A 2D superarray formed by two types of Rydberg atoms (blue and black sphere) trapped in optical tweezers. An energy detuning $\Delta$ is applied for all Rydberg states $|1\rangle$ on B sublattice. (b) Effective hopping processes produced by resonant exchange couplings (black arrows) and LADDI (red arrows). (c) The Rydberg dipole-dipole interaction can induce spin-exchange coupling between sites of the same type. However between A and B sites, the bare spin-exchange coupling is suppressed and is further recovered by a two-photon Raman process $\Omega$ (red arrow).
(d)The realized lattice structure with blue (black) dots representing the lattice sites A (B) and solid (dashed) lines representing positive (negative) exchange coupling coefficients. Arrows mark the direction of complex exchange coupling. The LADDI generates $\pi$-flux in each square plaquette. }
\end{figure*}

\textcolor{blue}{\em Introduction.}--Topological order is an exotic quantum order beyond Landau paradigm, whose discovery galvanizes the extensive search for topological quantum matter~\cite{Wen}. 
Fractional quantum Hall effect (FQHE)~\cite{FQHE1,FQHE2} is one of the most fascinating example of topological order. Generically, to realize FQHE requires two basic ingredients, i.e. the fractionally filled Laudau levels (LLs) created by strong (pseudo)magnetic field, and the interactions between particles. Various methods have been devised to emulate magnetic fields in quantum simulators, including rotating ultracold quantum gases~\cite{wilkin1998,madison2000,wilkin2000,cooper2001,ho2001,paredes2001,schweikhard2004,zwierlein2005,fetter2009,gemelke2010,mukherjee2022}, engineering complex tunnelings in real space~\cite{sorensen2005a,lin2009a,jotzu2014a,aidelsburger2015,kennedy2015a,tai2017}, synthetic dimension~\cite{mancini2015,an2017,zhou2023b} and photonic systems~\cite{roushan2017,ozawa2019}, which open up the possibility of creating and manipulating FQHE in engineered quantum systems. Recently, the bosonic $1/2$-Laughlin state~\cite{Bosoniclaughlin1}, a hallmark of the FQHE, has been demonstrated in photonic~\cite{clark2020} and ultracold atomic systems in the few-body regime~\cite{leonard2023c}. Another family of systems that exhibit FQHE-like behavior are fractional Chern insulators (FCIs)~\cite{FCI8,FCI9,FCI10}, which are lattice analogues of FQH phase with~\cite{Cooper1,FCIn1,FCIn2,FCIn3,FCIn4,FCIn5} or without LLs~\cite{FQAH1, FQAH2, FQAH3, FQAH4, FQAH5, FQAH6, FQAH7, FQAH8, FQAH9}. The latter are also called fractional quantum anomalous Hall (FQAH) insulator, and can be achieved by replacing fractionally filled LLs by fractionally filled topological flat bands that have zero net flux. Simulating FQAH phases could deepen our understanding of topological order and potentially reveal new topological ordered phases that have no continuum counterparts. In particular, for FQAH phases in topological flat band with Chern number $|C|=1$, the many-body ground states are lattice versions of FQH states. The adiabatic connection between such FQAHs and the corresponding FQH states, the bosonic $1/2$-Laughlin state, can be established by relating single-particle states in a $|C|=1$ flat band to those in a LL~\cite{FQAH1, mapping1, mapping2, mapping3, mapping4,  mapping5, mapping6}. Despite the similarity of the ground states in this scenario, system with FQAH potentially has larger band gap. The reason is that simulating FQH requires that the single particle band to be sufficiently flat, which can be achieved by choosing a sufficiently large $q$ in Hofstadter model~\cite{aidelsburger2015, kennedy2015a, an2017, tai2017, zhou2023b, leonard2023c} (with flux $p/q$), but such choice limits the size of the band gap. This poses a main challenge for experimentally engineering the ground states in the many-particle regime. The FQAH, including our proposal in this work, can be achieved on top of a minimal two-band model which has both large band gap and large band flatness. Furthermore, for FQAHs phases in flat bands with higher Chern number $|C|>1$, the lattice effects enrich the FQAH phases~\cite{FQAH8,sterdyniak2013,bergholtz2015,behrmann2016} and make them distinct from the conventional FQH phases described by Laughlin wavefunctions. These phases are not fully understood and offer rich prospects for both theoretical and experimental research.

The realization of FQAH phases is a formidable challenge, as it demands the simulation of topological flat bands with zero net flux, which are achieved by engineering exotic long-range hopping processes rather than applying (pseudo)magnetic fields. This entails the careful manipulation of long-range hopping processes. To realize such exotic phase has driven a lot of theoretical and experimental efforts -- The toy models have been proposed for realizing topological flat band~\cite{TFB1,TFB2,TFB3,Yao1}, and lots of attempts at realizing FQAH phases have been made, including in ultracold atoms~\cite{Yao2,Miao2022} and recently the electronic twisted bilayer materials~\cite{TBG0,TBG1,TBG2,TBG-FQHE-1,TBG-FQHE-2}, in which important progresses have been reported.

Recently, the Rydberg atoms, as characterized by highly excited atomic states, have shown the marvelous versatility in exploring intriguing correlated quantum matter~\cite{Browaeys2020}. Through the individual trapping by optical tweezers in a programmable array~\cite{Rydberg-tweezer1,Rydberg-tweezer2,Rydberg-tweezer3,Rydberg-tweezer4,Rydberg-tweezer5,Rydberg-tweezer6}, the Rydberg atoms with the long-range dipole-exchange interactions can naturally simulate hard-core bosons with the effective hopping between the array sites~\cite{Weber2018,Rydberg-lattice,Lienhard2020,Ohler2022}. The high tunability enables plenty of strongly correlated phenomena being observed therein, including quantum magnetism~\cite{Rydberg-spin1,Rydberg-spin2,Rydberg-spin3,Rydberg-spin4}, the three-site density-dependent Peierls phase~\cite{Lienhard2020}, the one-dimensional bosonic symmetry protected topological phase~\cite{Rydberg-lattice} and the two-dimensional quantum spin liquid~\cite{Rydberg-QSL}. With these marked progresses the interests in realizing bosonic FQAH have been recently revived in Rydberg atoms, and interesting schemes have been proposed recently~\cite{Weber2022,zhao2022}.
The neutral atoms may provide a highly controllable platform to precisely explore this strongly correlated topological order, whereas the experimental realization of the bosonic FQAH, being a challenging task, calls for further great efforts.

In this article, we propose a novel stripe-lattice model for realizing topological flat band generated via laser-assisted dipole-dipole interaction (LADDI)~\cite{LADDI,LADDI1} in a {\em supperarray} of  Rydberg atoms, and further propose the detection of the $\nu=1/2$ FQAH phase. The proposed superarray structure is formed by two different Rydberg states, between which the dipole interactions are assisted by Raman potentials, giving rise to a minimal two-band QAH model. The well-controlled Raman potentials, which have been actively studied theoretically and experimentally in optical lattice~\cite{ORL1,ORL2,ORL3,ORL4,ORL5}, are extended to the present Rydberg superarray to manipulate magnitude and relative phases of the LADDIs, offering a high tunability of the QAH model, which shows important advantages in realizing the FQAH phase. In particular, in combination with long-range dipolar-exchange interactions of Rydberg atoms, our Raman superarray scheme yields a robust topological flat band with large topological gap. Further, the natural open boundary of Raman superarray enables us to readily probe fractional excitation in real space. Experimental signatures including fractionalized accumulated charge and dynamics of fractional quasi-particle tunneling are predicted and examined carefully.

\textcolor{blue}{\em Model.}--We propose a strip-lattice model consisting of two types of Rydberg atoms situated in alternating rows (stripe structure) and trapped in a two dimensional optical tweezers coupled by LADDI (see \textbf{Figure~\ref{fig:SchemeFig}}), which has essential advantage in the realization of FQAH phase. The Hamiltonian $H$ reads
\begin{align}
\label{modelHAll}
H= & \text{\ensuremath{\sum_{\substack{j_y\in\mathrm{odd}\\l_y\in\mathrm{odd}}}\sum_{j_x,l_x}J_{l_x,l_y}a_{j_x,j_y}^{\dagger}b_{j_x+l_x,j_y+l_y}+\mathrm{h.c.}}} \nonumber \\
& +\sum_{\substack{j_y\in\mathrm{odd}\\l_y\in\mathrm{even}}}\sum_{j_x,l_x}J_{l_x,l_y}a_{j_x,j_y}^{\dagger}a_{j_x+l_x,j_y+l_y}\nonumber \\
 & +\sum_{\substack{j_y\in\mathrm{even}\\l_y\in\mathrm{even}}}\sum_{j_x,l_x}J_{l_x,l_y}b_{j_x,j_y}^{\dagger}b_{j_x+l_x,j_y+l_y},
\end{align}
where $a_{j_x,j_y}^{\dagger}$ ($b_{j_x,j_y}^{\dagger}$) creates a hard-core boson at site
$(j_x,j_y)$ with odd (even) $j_y$, simulated by the Rydberg states of type $A$ ($B)$. The hopping coefficients $J_{l_x,l_y}$ are independent of position except it picks up a phase of $e^{i\pi j_x l_y}$~\cite{LADDI,LADDI2,LADDI3}.

To give an intuitive example and later demonstrate high feasibility of realizing topological band with large flatness ratio, we consider first a minimal model including hoppings with $l_x^2+l_y^2\leq4$. The flatness ratio $E_\text{gap}/w$ is defined by the ratio of the band gap $E_{\mathrm{gap}}$ and the bandwidth of the lower band $w$. Explicitly, as illustrated in Figure~\ref{fig:SchemeFig}(a,b), the Hamiltonian that keeps terms up to next next nearest neighbor (NNNN) hopping is given by
\begin{eqnarray}
\label{modelH}
H&=& \sum_{\langle \boldsymbol{i},\boldsymbol{j}\rangle}J_{\boldsymbol{j}-\boldsymbol{i}}^{(1)}[(a_{\boldsymbol{i}}^{\dagger}a_{\boldsymbol{j}} +b_{\boldsymbol{i}}^{\dagger}b_{\boldsymbol{j}})+a_{\boldsymbol{i}}^{\dagger}b_{\boldsymbol{j}}+\mathrm{h.c.}] \nonumber\\
&& +\sum_{\langle\langle \boldsymbol{i},\boldsymbol{j}\rangle\rangle}(J_{\boldsymbol{j}-\boldsymbol{i}}^{(2)}a_{\boldsymbol{i}}^{\dagger}b_{\boldsymbol{j}}+\mathrm{h.c.})+\sum_{\langle\langle\langle \boldsymbol{i},\boldsymbol{j}\rangle\rangle\rangle}J_{\boldsymbol{j}-\boldsymbol{i}}^{(3)}(a_{\boldsymbol{i}}^{\dagger}a_{\boldsymbol{j}} +b_{\boldsymbol{i}}^{\dagger}b_{\boldsymbol{j}})\nonumber\\,
\end{eqnarray}
with $J_{\hat{\boldsymbol{e}}_x(\hat{\boldsymbol{e}}_y)}^{(1)}=J_{x(y)}$ being NN coefficient along $\hat{\boldsymbol{e}}_x(\hat{\boldsymbol{e}}_y)$-direction, $J_{\pm\hat{\boldsymbol{e}}_x+\hat{\boldsymbol{e}}_y}^{(2)}=J_{d1(d2)}$ being NNN coefficient along $\pm\hat{\boldsymbol{e}}_x+\hat{\boldsymbol{e}}_y$-direction, and $J_{2\hat{\boldsymbol{e}}_x(2\hat{\boldsymbol{e}}_y)}^{(3)}=J_{\mathrm{nn}x}(J_{\mathrm{nn}y})$ being NNNN hopping along $\hat{\boldsymbol{e}}_x(\hat{\boldsymbol{e}}_y)$-direction, respectively. Identical to Eq.~\eqref{modelHAll}, the hopping coefficients $J_y$, $J_{d1}$ and $J_{d2}$ pick up a $\pi$-phase of $e^{i\pi j_y}$, manifesting itself the stripe sublattice structure [Figure~\ref{fig:SchemeFig}(b,d)]. This stripe-lattice model that we propose in Eq.~\eqref{modelHAll} and Eq.~\eqref{modelH} is quite different from the previously proposed checkerboard-lattice models~\cite{FQAH3,FQAH4,TFB2,TFB3,Yao1,Yao2}. In the following context, we show that the minimal model~\eqref{modelH} can realize a topological flat band for hard core boson with $E_\text{gap}/w \approx 22$. Further including longer range hoppings and their higher order corrections from Raman lights remarkably enhances the flatness to $E_\text{gap}/w \approx 38$, facilitating the realization of the FQAH states with Raman superarrays.

\textcolor{blue}{\em Experimental Realization.}--Here, we propose an experimentally accessible and tunable scheme that coupled two types of Rydberg states by LADDI to achieve the above model. More specifically, we propose to realize our effective Hamiltonian (\ref{modelHAll}) through
\begin{align}
H_m =& \sum_{\vec{r}_1,\vec{r}_2} J^{\text{DDI}}_{\vec{r}_1\rightarrow \vec{r}_2}\sigma^{+}_{\vec{r}_2}\sigma^{-}_{\vec{r}_1} + \mathrm{h.c.} \nonumber \\
&+ \frac{\Delta}{2}\sum_{\substack{\vec{r}_{m,n}\\n\in\text{even}}}\sigma^{z}_{\vec{r}}+\sum_{\vec{r}}\Omega(\vec{r})e^{i\delta\omega t}\sigma^{x}_{\vec{r}},
\end{align}  
where operator $\sigma^{+}_{\vec{r}_{m,n}} = \left|1,\vec{r}_{m,n}\right>\left<0,\vec{r}_{m,n}\right|$, with $\vec{r}_{m,n} = m \vec{a}_1+ n \vec{a}_2 = m(a_1,0) + n a_2 (\cos\theta,\sin\theta)$ being the location of the Rydberg atom, $\theta$ being the angle between two lattice vectors $\vec{a}_1$ and $\vec{a}_2$ [Figure~\ref{fig:SchemeFig}(a)]. Two Rydberg states of $^{87}\mathrm{Rb}$ atoms $\left|0\right> = \left|60S_{1/2}, m_j = 1/2\right>$ and $\left|1\right> = \left|60P_{1/2}, m_j=1/2\right>$ are chosen to simulate two-level system at each site and are quantized along the axis $\vec{q} = \left(\sin\theta_q \cos\phi_q, \sin\theta_q\cos\phi_q,\cos\theta_q\right)$. A detuning $\Delta$ is applied for all Rydberg states $\left|1\right>$ in $B$ sublattice~\cite{detuneNote}. The dipole-dipole interaction (DDI) between two Rydberg states leads to an exchange coupling, which can be mapped to hopping of hard-core bosons. Therefore, as illustrated in first two panels Figure~\ref{fig:SchemeFig}(c), the hoppings between Rydberg states of the same type -- $J_x,$ $J_{\mathrm{nn}x}$ and $J_{\mathrm{nn}y}$ -- are produced by bare DDI between $\left|0\right>$ and $\left|1\right>$. The key ingredient of the scheme is that the bare DDI between Rydberg states of the different type -- $J_y^0$ , $J_{d1}^0$ and $J_{d2}^0$ -- are suppressed by the relatively large energy penalty $\Delta$, but can be restored by Raman potential $\Omega(r)$
with frequency difference $\delta\omega \approx\Delta$ to compensate the energy detuning $\Delta$, rendering two Rydberg states coupled by LADDI [see Figure~\ref{fig:SchemeFig}(c)]. 

We now show the quantitative results of the hopping coefficients induced by LADDI for two alkali Rydberg atoms at $\vec{r}_1$ and $\vec{r}_2$, respectively. For LADDI, the hopping, up to leading order, is written as (derivation and more accurate expression that includes infinite order are shown in appendix~\cite{SM})
\begin{equation}
J^{\text{LADDI}}_{\vec{r}_1\rightarrow \vec{r}_2} = \frac{C_3(\theta_{\vec{r}_2-\vec{r}_1})}{ r^3}\frac{\left[\Omega(\vec{r}_1) - \Omega(\vec{r}_2)\right]^{(*)}}{\delta \omega},
\label{LADDI}
\end{equation}
where $C_3(\theta_{\vec{r}_2-\vec{r}_1})=C_3\left(1-3\cos^2(\theta_{\vec{r}_2-\vec{r}_1})\right)$, with $C_3 = -3000 \text{MHz} \cdot \mu m^3$ \cite{ARC} is the matrix element of the DDI between $\left|0\right>$ and $\left|1\right>$, $\theta_{\vec{r}_2-\vec{r}_1}$ is the angle between $\vec{r}_1 - \vec{r}_2$ and $\vec{q}$, $r$ is the distance between the two atoms, and $\Omega (\vec{r}_i)$ is the Raman potential at the position $\vec{r}_i$.  The conjugate $(*)$ is needed to be applied when $\vec{r}_2$ is of detuning $\Delta$ and $\vec{r}_1$ is of no detuning; whereas no conjugate is applied when $\vec{r}_2$ is of no detuning and $\vec{r}_1$ is of detuning $\Delta$. The term $J_y$ is produced by LADDI with $\Omega_1 = \left|\Omega_1\right| e^{i\vec{k}_1 \cdot \vec{r}}/2$, where $\vec{k}_1 = \frac{\pi}{a_2} \left(\alpha, \frac{1-\alpha\cos\theta}{\sin\theta}\right)$, so that $\Omega_1=\pm \left|\Omega_1\right|/2$ for even or odd $m+n$ respectively. Therefore, $J_y = e^{i\pi j_x} a_2^{-3} C_3(\theta_{\vec{a}_2})  \linebreak[1] \left|\Omega_1\right|$.~The terms $J_{d1}$ and $J_{d2}$ are produced by LADDI with $\Omega_2 = \left|\Omega_2\right|e^{i(\vec{k}_2 \cdot \vec{r}+\pi/2)}/2$, where $\vec{k}_2 = \frac{\pi}{a_1}\left(1, -\cot\theta\right)$, so that $\Omega_2=\pm i\left|\Omega_2\right|/2$ for even or odd $m$ respectively. Then $J_{d1(d2)} = ie^{i\pi j_x}a_1 \linebreak[1] \left|(1\pm\alpha \sin\theta,\alpha\cos\theta)\right|^{-3/2} \linebreak[1] C_3(\theta_{\vec{a}_1\pm \vec{a}_2}) \left|\Omega_2\right|$. The Raman potential $\Omega_1$ does not produce LADDI between atoms with $\left|\Delta m\right| = \left| \Delta n \right| = 1$ since $\Omega_1$ takes the same value at $r_1$ and $r_2$ [Equation (\ref{LADDI})]. Similarly, $\Omega_2$ does not produce LADDI between atoms with $\left|\Delta n\right| = 1$.

In addition, the bare DDI is also dressed by the Raman potential applied. The hopping between two atoms of the same type is written as (details in appendix~\cite{SM})
\begin{equation}
J^{\text{DDI}}_{\vec{r}_1\rightarrow \vec{r}_2} = \frac{C_3(\theta_{\vec{r}_2-\vec{r}_1})}{r^3}c(\Omega),
\label{bddi}
\end{equation}
where $c(\Omega)= 1+\left[\Omega(\vec{r}_1)-\Omega(\vec{r}_2)\right] \left[\Omega^*(\vec{r}_2)-\Omega^*(\vec{r}_1)\right]\delta\omega^{-2}  = \left(1-4\Omega^2\delta\omega^{-2}\sin^2 \phi/2 \right)$ is the correction, up to second order, due to the Raman potential, and $\omega$ is the frequency of the Raman potential. The last equality holds when $\left|\Omega(\vec{r}_1)\right| = \left|\Omega(\vec{r}_2)\right|$ and $\phi$ is the phase differences between $\Omega (\vec{r}_1)$ and $\Omega (\vec{r}_2)$. According to (\ref{bddi}), $J_x = C_3(\theta_{\vec{a}_1}) a_1^{-3} (1-4(\left|\Omega_1\right|^2+\left|\Omega_2\right|^2)/\omega^2)$, $J_{\mathrm{nn}x} = C_3(\theta_{\vec{a}_1}) (2a_1)^{-3}$ and $J_{\mathrm{nn}y} = C_3(\theta_{\vec{a}_2}) (2a_2)^{-3}$. Note here that $J_{\mathrm{nn}x}$ and $J_{\mathrm{nn}y}$ have no correction since both Raman potentials have the same value at two sites with $(\Delta m,\Delta n) = (\pm2,0)$ or $(0,\pm 2)$ so that $\sin^2 \phi/2 = 0$. Summing up all the above couplings and performing a gauge transformation $\sigma^{+}_{j_x,j_y} \rightarrow \cos(\pi j_x)\sigma^{+}_{j_x,j_y}$ and renaming the operators at even (odd) site as $A$ ($B$), we reach the Hamiltonian \eqref{modelH}.

\begin{figure*}[tp]
\begin{center}
\includegraphics{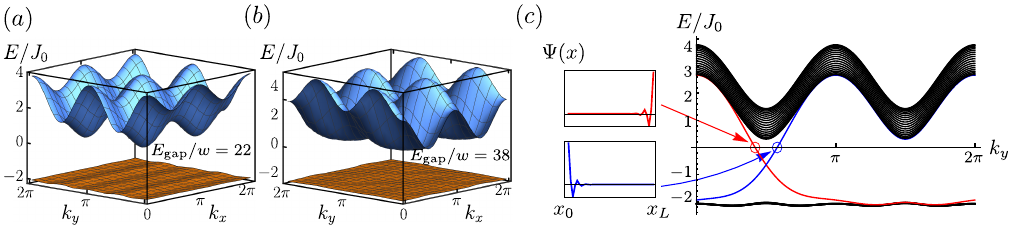}
\end{center}

\caption{\label{fig:SingleFig}The two-band spectrum of the realized QAH model and the flatness ratio. (a) The spectrum and the flatness ratio $E_\text{gap}/w$  when the experimental parameters are tuned to optimal for the minimal case $l_x^2+l_y^2\leq 4$. (b) Similar diagram for the case $l_x^2+l_y^2\leq 9$. (c) The spectrum of the minimal model on a cylinder geometry and corresponding topological edge states.}
\end{figure*}

\textcolor{blue}{\em Topological flat band.}--The above scheme is highly tunable and by choosing suitable parameters, a topological flat band can be obtained feasibly. The system is controlled by six fundamental parameters -- (i,ii) The angle $\theta$ and the ratio $\alpha=a_2/a_1$ determine the structure of the Raman superarray and can be adjusted flexibly. (iii,iv) The angles $\theta_q$ and $\phi_q$ that define the quantization axis can be readily tuned by the direction of the external magnetic field. (v,vi) The Raman potentials $|\Omega_{1(2)}|$ are controlled by the strength of the Raman light. (See appendix \cite{SM} for more details about the experimental feasibility). The optimal parameters to reach topological flat band can be numerically found by simulated annealing. For the minimal case $l_x^2+l_y^2\leq4$, by setting the optimal parameters $\left|\Omega_1\right| = 0.5\omega$, $\left|\Omega_2\right| = 0.442\omega$, $\alpha=0.657$, $\theta = 68.5^{\circ}$, $\theta_q = 74^{\circ}$ and $\phi_q = 29.3^{\circ}$, a nontrivial topological flat band with band gap $E_\text{gap}=2.36J_0$ and $E_\text{gap}/w\approx22$ is obtained as shown in \textbf{Figure~\ref{fig:SingleFig}}(a,c). The dispersion relations in \ref{fig:SingleFig}(a,b) are obtained on periodic boundary conditions and the spectrum in \ref{fig:SingleFig}(c) is obtained on a cylinder with open (periodic) boundary condition along $x$ ($y$) direction. The quantity $J_0 = C_3 a_1^3$ defines a typical energy scale of the system. The corresponding hopping coefficients are $J_{x}=-0.87J_0$, $J_y=1.18J_0$, $J_{\mathrm{nn}x}=0.14J_0$, $J_{\mathrm{nn}y}=0.29J_0$ and $J_{d_{1}}=-J_{d_{2}}=i|J_d|=0.29iJ_0$ in this regime. A remarkable feature of our scheme is that $f$ can be drastically improved when terms with longer range hopping are taken into consideration. A system with topological flat band exhibits a suppression of particle hopping between adjacent unit cells, resulted from destructive interference of the hopping through different paths. By the presence of longer range hopping, the interference effects are enhanced due to the increased number of interfering paths, which further flattens the topological flat band. In particular, we include all hopping terms satisfying $l_x^2 + l_y^2 \leq 9$ and found that at optimal parameters $\left|\Omega_1\right| = 0.482\omega$, $\left|\Omega_2\right| = 0.654\omega$, $\alpha=0.874$, $\theta = 124.6^{\circ}$, $\theta_q = 152.1^{\circ}$ and $\phi_q = 66.1^{\circ}$~\cite{SM}, $E_\text{gap}/w$ is greatly improved to $E_\text{gap}/w \approx 38$ with gap $E_\text{gap} = 1.67J_0$ [Figure~\ref{fig:SingleFig}(b)]. A similar improvement of flatness by longer range hopping was studied in the Hofstadter model in optical lattice for FQH states~\cite{Kapit2010}, and our work provides an experimental scheme to manipulate such hoppings in a novel stripe-lattice model to maximize the flatness ratio and topological band gap, enabling a feasbile simulation of FQAH phase with Rydberg atoms.

\begin{figure*}[tp]
\begin{center}
\includegraphics{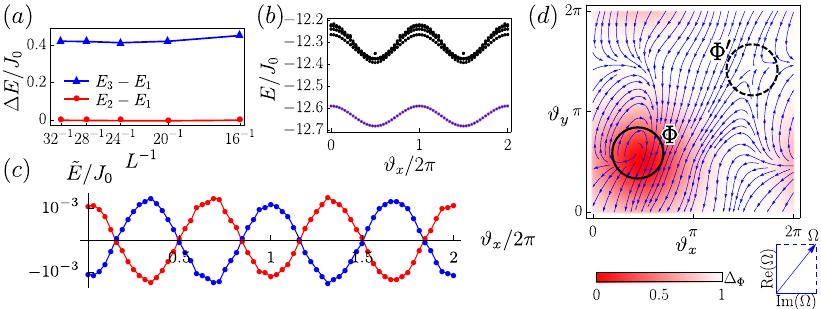}
\end{center}

\caption{\label{fig:ED Fig}Features of $\nu=1/2$ bosonic FQAH state. (a)
The finite-size scaling for both ground state splitting and
energy gap for $N=4$, $5$, $6$, $7$ and $8$, $L_x=N$, $L_y=4$. The results show $2$-fold degeneracy of the many-body
ground states, with a finite gap separating from the excited states.
(b) The spectral flow for $L_{x}\times L_{y}=4\times6$, where $\vartheta_{x}$
is the magnetic flux insertion along $x$-direction. Evolution of
2-fold degenerate ground states upon flux insertion. The two ground states maintain their quasidegeneracy in the presence of flux insertion and well separated from other low excited states. (c) The spectral flow of two ground states shows a characteristic swap within $2\pi$ pumping. Here $\tilde{E} = E_{1(2)}-(E_1 + E_2)/2$. (d) At the place where $\Delta_\Phi = 0$ (solid circle), $\Omega(\vartheta_x,\vartheta_y)$ field has vorticity equal to 1, indicating two degenerate multiplets share unit Chern number, so $C=1/2$ for a single ground state~\cite{SM}. Equivalently, the vorticity of $\Omega(\vartheta_x,\vartheta_y)$ at the place where $\Delta_{\Phi'} = 0$ (dashed circle) also indicates the non-trivial topology of the two degenerate multiplets. Here $N=4$ and $L_x\times L_y=4\times6$.}
\end{figure*}

\textcolor{blue}{\em  Bosonic 1/2 FQAH phase.}--With the topological flat band realized, we now show three characteristic features of the many body ground state at filling  $\nu=1/2$ (number of particles per unit cell) that epitomize bosonic $1/2$-FQAH phase -- (i) degenerate ground states with finite spectral gap, (ii) spectral flow under twisted boundary condition and (iii) the many-body Chern number (MBCN) $\emph{C}$. These features are consistent with the bosonic $1/2$-Laughlin state~\cite{Bosoniclaughlin1}, as the FQAH states and the corresponding FQH states can be adiabatically mapped to each other by relating the single-particle states in a topological flat band to those in a Landau level~\cite{FQAH1,mapping1,mapping2,mapping3,mapping4,mapping5,mapping6}. Firstly, there is a two-fold quasi-degenerate ground states well separated from other excited states by a pronounced gap on a torus. Here, we exactly diagonalize the Hamiltonian [Eq.~\eqref{modelHAll}] for $N$ particles in a $L_x\times L_y$ lattice, where we have $L_x\times L_y=4N$. We calculate the excitation gaps for $L_x=4$, $5$, $6$, $7$ and $8$, $L_y=4$. The gap is large compared to the ground state splitting $E_{3}-E_{2}\gg E_{2}-E_{1}$ and is very likely to persist in the thermodynamic limit as indicated by the finite-size scaling analysis shown in \textbf{Figure~\ref{fig:ED Fig}}(a). Secondly, the two-fold ground states remain quasi-degenerate in twisted boundary condition and show a 4$\pi$-periodicity of spectral flow. For a many-body state $|\Psi\rangle$,
the twisted boundary condition in the $x$($y$)-direction is $\langle\boldsymbol{r}+N_{x}\hat{\boldsymbol{e}}_x(N_{y}\hat{\boldsymbol{e}}_y)|\Psi\rangle=e^{i\vartheta_{x(y)}}\langle\boldsymbol{r}|\Psi\rangle$,
where $\vartheta_{x(y)}$ is the twisting angle. As $\vartheta_x$ changes from 0 to $4\pi$, the two ground states maintain their quasi-degeneracy and excitation gaps [Figure~\ref{fig:ED Fig}(b)], evolve into each other and finally move back to the initial configuration [Figure~\ref{fig:ED Fig}(c)]. The spectral flow shows the presence of $\nu=1/2$ fractional quantum Hall state. Thirdly, the MBCN~\cite{ChernNum,TBC} of the ground state is found to be $C=1/2$. The calculation starts by defining the ground state multiplet $\Psi_g(\vartheta_x,\vartheta_y) = (|\Psi^{(1)}\rangle, |\Psi^{(2)}\rangle)$ with $|\Psi^{(1,2)}\rangle$ being the two degenerate ground states for the system with twisted boundary conditions in both directions, two multiplets $\Phi=\Psi_g(0,0)$ and $\Phi'=\Psi_g(\pi,\pi)$  for gauge references, the overlapping matrix $\Delta_\Phi(\vartheta_x,\vartheta_y) = \det (\Phi^{\dagger} \Psi_g \Psi_g^\dagger \Phi)$. Then the MBCN for the two-fold degenerate ground states can be obtained by counting the vorticity of $\Omega(\vartheta_x,\vartheta_y) = \det (\Phi^{\dagger} \Psi_g \Psi_g^\dagger \Phi')$ where $\Delta_\Phi$ (or $\Delta_{\Phi'}$) vanishes \cite{SM,MBCN1,MBCN2}. As illustrated in Figure~\ref{fig:ED Fig}(d), two degenerate ground states share unit MBCN, corresponding to a ground state with $C=1/2$. These results unambiguously determine the existence of the bosonic $1/2$-FQAH phase.

\begin{figure*}[tp]
\begin{center}
\includegraphics{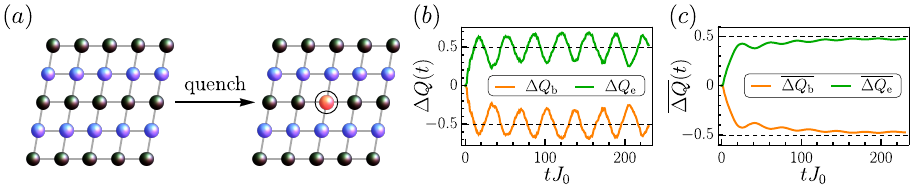}
\end{center}

\caption{\label{fig:SingleDefect} Detection of fractionalized particles via correlated quench dynamics. (a) Quench protocol. The blue and black spheres represent different types of Rydberg states. Single repulsive potential $+V$ (red sphere) is introduced at the center of system with open boundary. The initial state is prepared in a clean FQAH state (left) and is then quenched to system with impurity (right). (b) Quench dynamics of net charge $\Delta Q_{e(b)}(t)$ of the edge (bulk) on a $L_x\times L_y = 5\times5$ superarray with $N=4$ and $V=3.0J_0$. (c) The time averaged net charge $\overline{\Delta Q}(t)$ versus time $tJ_0$.}
\end{figure*}

\textcolor{blue}{\em Detection protocol.}--Finally we propose a novel quench protocol to detect fractional excitation in real space [\textbf{Figure~\ref{fig:SingleDefect}}(a)]. For a $\nu=1/2$ FQAH phase, the edge of the system serves as a reservoir of fractionlized quasi-particles and quasi-holes. One noteworthy feature of Rydberg system is its intrinsic sharp open boundary, unlike other simulation platforms such as ultracold atoms, which require additional manipulations to create sharp open boundaries~\cite{Braun2023,yao2023}.
Therefore, we can pinpoint the region of edge precisely in the system and utilize this advantage to probe fractionalized quasi-particle in real space. We present the supplementary static properties of fractionalized excitation in real space, calculated by density matrix renormalization group (DMRG)~\cite{schollwock2011density,White1992,fishman2021} in the appendix. In this section, we focus on the non-equilibrium quench protocol. The key idea is to trap quasi-particles (-holes) by introducing attractive (repulsive) potentials in the post-quench Hamiltonian and measure the change in population of Rydberg states. Such manipulation is readily accessible to Rydberg experiment by introducing extra energy detuing $V$ through AC stark shift on Rydberg state $|1\rangle$ at the corresponding positions as illustrated in Figure~\ref{fig:SingleDefect}(a)(right), and the measurement is also accessible through fluorescence imaging. The net charge $\Delta Q_R$, the evolution of population of ground state $|\Psi_V\rangle$ from $|\Psi_{V=0}\rangle$ in a given region $R$, is defined as
\begin{equation}
\Delta Q_{R}=
\sum_{j\in R}\left[\langle \Psi_{V} \left|1\rangle_j \langle 1 \right|_j\Psi_{V}\rangle-\langle \Psi_{V=0} \left|1\rangle_j \langle 1 \right|_j\Psi_{V=0}\rangle \right].
\end{equation}
We probe $\Delta Q_R$ from the correlated quench dynamics with open boundary conditions. More specifically, we prepare clean FQAH state with $V=0$ as initial state $\lvert \Psi(t=0) \rangle$ and suddenly quench it to the system with repulsive potential at center with $V=3.0J_0$, which corresponds to the post-quench Hamiltonian $H_{\mathrm{post}}$. We study the system of size $L_x=L_y=5$ and particle number $N=4$, and obtain the post-quench state by exactly evolving the initial state $\lvert \Psi(t=0) \rangle$ under the $H_{\mathrm{post}}$ Hamiltonian, resulting in $\lvert \Psi(t) \rangle = \exp (i H_{\mathrm{post}}t)\lvert \Psi(t=0) \rangle$. The net charge $\Delta Q_{\mathrm{e}(\mathrm{b})}(t)$ of edge (bulk) coherently oscillates around $\pm1/2$ and the time average net charge $\overline{\Delta Q_{\mathrm{e}(\mathrm{b})}}(t) = t^{-1} \int_0^t d\tau \Delta Q(\tau)$ is quantized to $\pm1/2$ [Figure~\ref{fig:SingleDefect}(b,c)], indicating the many-body dynamics featured by tunneling of the fractionlized quasi-particle and quasi-hole. The net charge $\Delta Q_{\mathrm{e}(\mathrm{b})}$ of the equilibrium state, measured from ground state of the post-quench Hamiltonian, are shown in the supplementary material~\cite{SM}. This correlated dynamical response simultaneously detects the fractionalized excitations in the bulk and the chiral edge, in comparison with the equilibrium measurement of fractional excitations~\cite{realspace1, realspace2, realspace3, realspace4, realspace5, realspace6}.

\textcolor{blue}{\em Conclusions and Discussions.}--We have proposed a novel scheme dubbed {\em Raman superarrays of Rydberg atoms} to simulate a strongly correlated stripe-lattice model for hard-core bosons with high flatness ratio $E_\text{gap}/w$, from which we propose to realize and detect a $\nu=1/2$ bosonic fractional QAH phase. In particular, we considered two types of Rydberg states arranged in a stripe configuration, and the intrinsic DDI between them is suppressed but can be assisted by Raman coupling (i.e. the LADDI), which realizes a correlated QAH model. By adjusting the experimental parameters, the lower Chern band can exhibit a high flatness, which is essentially enabled by the presence of long range couplings in the LADDI. Such stripe lattice configuration can be achieved through various methods with high tunability, including AC stark shift, which we considered in the main text for simpler discussion, using pairs of states with different principal quantum number, or using atoms of different species~\cite{detuneNote}. Finally, we proposed a novel quench protocol to simultaneously detect the bulk fractional excitations and the chiral edge fractional particles. By suddenly tuning an onsite potential in the bulk, the correlated many-body quench dynamics have been predicted and featured by the tunneling of the $1/2$-fractionalized particles between the bulk defect and the edge. This provides a clear signature for the FQAH phase realized for the present Rydberg superarrays. Our scheme can be readily extended to realize the topological flat band with higher Chern number $|C|$>1, which can host more exotic FQAH phases that have no continuum analogue. Recent studies have demonstrated the possibility of realizing FQH states, in the platform of ultracold atom and photonic systems, but the realization of its zero-net-flux counterpart FQAH generically demands the tuning of longer range hopping, which is challenging in these platforms, to engineer a topological flat band. The platform of Rydberg atom inherently possesses longer range coupling and our scheme provides a feasible way to adjust the required coupling. Thus, this work may prove successful in the pursuit of simulating such exotic physics in simulation platforms, and may open a broad avenue to simulate strongly correlated system in the platform of Rydberg atoms.

\textcolor{blue}{\em Acknowledgments.}--The parameters of Rydberg atoms are obtained from "ARC" package~\cite{ARC} and the DMRG algorithm is implemented based on the ITensor library~\cite{fishman2021}. This work was supported by National Key Research and Development Program of China (2021YFA1400900),  the Innovation Program for Quantum Science and Technology (Grant No. 2021ZD0302000), the National Natural Science Foundation of China (Grants No. 11825401, No. 12261160368, and No. 11921005), and the Strategic Priority Research Program of the Chinese Academy of Science (Grant No. XDB28000000).

\section*{Appendix A: Single particle Properties}

The lower (upper) band of the minimal model [Eq.~\eqref{modelH}] is topological and processes a non-trivial Chern number $\pm 1$. To see this, we write the Hamiltonian in the momentum space as
\begin{align*}
H(k_x,k_y) =& h_0\sigma_0 + \vec{h} \cdot \vec{\sigma} \nonumber \\
=& \left(2J_{nnx}\cos 2k_x + 2J_{nny}\cos 2k_y\right)\sigma_0 \nonumber \\
&+ 2J_x\cos k_x \sigma_x + 4|J_{d1}| \sin k_x\sin k_y\sigma_y \nonumber \\
&+ 2J_y\cos k_y\sigma_z,
\end{align*}
where we have assumed $J_{d1} = -J_{d2}$ to be purely imaginary, $J_x$ and $J_y$ are real and that all $J$'s are non-zero. Along the band inversion line, which is the root of $h_z=0$, i.e. $k_y = \pm \pi/2$, $h_x+i h_y = 2J_x \cos k_x \pm 4i|J_{d1}|\sin k_x$ winds about $0$ whenever $J_x$ and $|J_{d1}|$ is nonzero so that the bands are topological~\cite{FHChernNumCount,BIS}. If we consider all hoppings up to NNNN, the parameters $J_x = -0.87J_0, J_y= 1.18J_0, J_{d1} = -J_{d2} =  0.29iJ_0, J_{nnx}=  0.14J_0$ and $J_{nny}= 0.29J_0$ are accessible in our proposal such that the flatness becomes $E_\text{gap}/w \approx 22$, and the gap $E_\text{gap} = 2.36 J_0$. On the other hand, if we consider all hoppings satisfying $l_x^2+l_y^2\leq 9$ and higher order corrections of the hopping due to Raman lights, an accessible and optimal set of parameters are shown in \textbf{Table~\ref{Tb: N^7parameters}}, with which the flatness of the system becomes $E_\text{gap}/w \approx 38$, and the gap $E_\text{gap} = 1.67 J_0$. We emphasis here that both sets of parameters are accessible by tuning the same six fundamental parameters (main text).

\begin{table}[hbp]
\begin{center}
\begin{tabular}{|c|c|c|c|c|c|}
\hline
hopping coefficients & $J_{x}$ & $J_{y}$ & $J_{\mathrm{nn}x}$ & $J_{\mathrm{nn}y}$ & $J_{d1}$ \tabularnewline
\hline
values/$J_{0}$ & $0.92$ & $0.193$ & $0.26$ & $-0.056$ & $-0.43i$ \tabularnewline
\hline
hopping coefficients & $J_{d2}$ & $J_{3,0}$ & $J_{2,1}$ & $J_{-2,1}$ & $J_{1,2}$ \tabularnewline
\hline
values/$J_{0}$ & $0.054i$ & $0.034$ & $-0.21$ & $-0.023$ & $-0.11$\tabularnewline
\hline
hopping coefficients & $J_{-1,2}$ & $J_{0,3}$ & $J_{2,2}$ & $J_{-2,2}$&\tabularnewline
\hline
values/$J_{0}$ & $0.003$ & $0.007$ & $-0.1$ & $0.013$&\tabularnewline
\hline
\end{tabular}
\end{center}
\caption{\label{Tb: N^7parameters}Values of the hopping coefficients for $l_x^2+l_y^2\leq 9$}
\end{table}
Here, we also demonstrate that the edge state is sufficiently thin so that the size of the system in our many-body calculation is valid. Such calculation also helps the experimental identification of the edge state. We consider the Hamiltonian with one direction being Fourier transformed
\begin{widetext}
\begin{eqnarray}
H(k_y) &=& \sum_{j_x\in\mathrm{even}} \left(e^{ik_y}J_y+e^{2ik_y}J_{nny}\right) \left(a^\dagger_{j_x,k_y} a_{j_x,k_y}-b^\dagger_{j_x-1,k_y} b_{j_x-1,k_y}\right) + J_{nnx} \left(a^\dagger_{j_x,k_y} a_{j_x+2,k_y}+b^\dagger_{j_x-1,k_y} b_{j_x+1,k_y}\right) \nonumber\\
&&+\left[\left(J_x + J_{d1}e^{ik_y}\right)a^\dagger_{j_x,k_y}b_{j_x+1,k_y} + \left(J_x + J_{d2}e^{ik_y}\right)a^\dagger_{j_x,k_y} b_{j_x-1,k_y}\right] + h.c. \\
H(k_x) &=& \sum_{j_y} e^{2ik_x} J_{nnx} \left(a^\dagger_{k_x,j_y} a_{k_x,j_y}+b^\dagger_{k_x,j_y} a_{k_x,j_y}\right)+e^{ik_x} J_x \left(a^\dagger_{k_x,j_y} b_{k_x,j_y}+b^\dagger_{k_x,j_y} a_{k_x,j_y}\right) \nonumber\\
&& + J_y \left(a^\dagger_{k_x\beta,j_y} a_{k_x\beta,j_y+1}-b^\dagger_{k_x\beta,j_y} b_{k_x\beta,j_y+1}\right) + J_{nny} \left(a^\dagger_{k_x\beta,j_y} a_{k_x\beta,j_y+2}+b^\dagger_{k_x\beta,j_y} b_{k_x\beta,j_y+2}\right)\nonumber\\
&& + \left(e^{ik_x} J_{d1}+e^{-ik_x} J_{d2}\right) \left(a^\dagger_{k_x\beta,j_y} b_{k_x\bar{\beta} j_y+1}-b^\dagger_{k_x\beta,j_y} a_{k_x\bar{\beta} j_y+1}\right) + h.c.
\end{eqnarray}
\end{widetext}
Note that the Hamiltonian $H(k_x)$ still contains the sublattice degree of freedom. The properties concerning the depth of the edge state can be captured through the effective Hamiltonian
\begin{equation}
H_{\text{eff}} = \left(
\begin{array}{cc}
I & 0\\
eI-H_0 & -B_0
\end{array}
\right) - \lambda
\left(
\begin{array}{cc}
0 & I\\
B_0^\dagger & 0
\end{array}
\right),
\end{equation}
where $e$ is the target energy of the edge state and $\lambda$ is the parameter in the trial wave function $e^{\lambda x}$ (or $e^{\lambda y}$)~\cite{edge}. For a particular $k_x$ or $k_y$, $H_0$ is the onsite Hamiltonian and $B_0$ is the hopping matrix. The eigenvalue $\lambda$ that describes the decay rate can be determined by the following formula
\begin{eqnarray}
\lambda_\gamma &=& -\frac{Q_{1\gamma}}{4} - \frac{1}{4}\sqrt{8+Q_{1\gamma}^2-4Q_{2\gamma}} - \frac{1}{2}\times \nonumber\\
&&\sqrt{-2+\frac{Q_{1\gamma}^2}{2}-Q_{2\gamma}-\frac{-8Q_{1\gamma}-Q_{1\gamma}^3+4Q_{1\gamma}Q_{2\gamma}}{2\sqrt{8+Q_{1\gamma}^2-4Q_{2\gamma}}}}\nonumber\\,
\end{eqnarray}
where
\begin{eqnarray}
Q_{1\gamma} &=& J_{nn\bar{\gamma}}^{-2}\left(2J_{d1} J_{d2} - 2 e J_{nn\gamma} - J_\gamma^2 + \right.\nonumber\\
&&\left.\left(J_{d1}^2 + J_{d2}^2 + 4 J_{nnx}J_{nny}\right)\cos 2 k_\gamma\right) \nonumber\\,
Q_{2\gamma} &=& J_{nn\bar{\gamma}}^{-2}\left(e^2+2J_{d1}^2+2J_{d2}^2+2J_{nnx}^2+2J_{nny}^2\right.\nonumber\\
&&\left.-2J_x^2-2J_y^2+2(2J_{d1}J_{d2}-2eJ_{nn\gamma}-J_\gamma^2)\cos 2k_\gamma\right.\nonumber\\
&&\left. + 2J_{nn\gamma}^2 \cos 4k_\gamma \right) \nonumber
\end{eqnarray}
,$\gamma = x$ or $y$ denotes the direction that we choose the periodic boundary condition and performed the Fourier transform, and $\bar{\gamma}$ denotes the orthogonal direction that we choose the open boundary condition. Typical $\lambda$'s are obtained at $e=0$, at which there is an edge state with $k_x = 1.322$ and $\lambda_x = -3.095$, or with $k_y = 1.295$ and $\lambda_y = -2.001$. For our many-body calculation, we choose $L_y = 5$ so that the overlapping of the edge states in opposite side is sufficiently small.

\section*{Appendix B: Derivation of the LADDI and the correction of bare DDI due to Raman process}
The correction to the bare DDI for two Rydberg atoms at $r_1$ and $r_2$ with same detuning can be captured by the Floquet Hamiltonian
\begin{equation}
H_T = \left(
\begin{array}{ccccc}
\ddots & \ddots & \ddots & &\\
\ddots & \mathcal{J}-\delta\omega & \mathcal{Q} & 0 &\\
\ddots & \mathcal{Q}^\dagger & \mathcal{J} & \mathcal{Q} & \ddots\\
& 0 & \mathcal{Q}^\dagger & \mathcal{J}+\delta\omega & \ddots\\
& & \ddots & \ddots & \ddots\\
\end{array}
\right),
\end{equation}
where $\mathcal{J} = \big(\begin{smallmatrix} 0 & J\\J & 0  \end{smallmatrix}\big)$, $\mathcal{Q} =  \big(\begin{smallmatrix} \Omega & 0\\0 & \Omega'   \end{smallmatrix}\big)$, $\delta\omega$ is the frequency of the Raman potential, $\Omega$ and $\Omega'$ is the strength of the Raman potential at $\vec{r}_1$ and $\vec{r}_2$, and $J$ is the strength of the bare DDI. The Floquet Hamiltonian $H_T$ can be transformed via the degenerate perturbation theory~\cite{DegenP} into an effective $2$-by-$2$ Hamiltonian $H_{\text{eff}}$ that correctly matches the eigenvalues and eigenstates of the full Hamiltonian.
For this purpose, we split the Hamiltonian into two parts $H_T = H_{0T} + V_T$, where $H_{0T}$ ($V_T$) is the (off-)diagonal part of $H_{T}$. Define $P$ to be the operator that project into the subspace with $E_0 = 0$, $Q = 1-P$, $V_{0M} = P V_T P$, $V_{iM} = P V_T Q$, $V_{fM} = Q V_T P$, $V_{cM} = Q V_T Q$ and $l = \left(E_0 - Q H_{0T} Q\right)^{-1}$. Then the effective Hamiltonian reads
\begin{align}
H_{\text{eff}}=& V_{0M} + V_{iM}lV_{fM} + \left(V_{iM}lV_{cM}lV_{fM}- \frac{1}{2}V_{iM}l^2V_{fM}V_{0M}-\right.\nonumber\\
&\left.\frac{1}{2}V_{0M}V_{iM}l^2V_{fM}\right)+ \left(\frac{1}{2}V_{iM}lV_{cM}lV_{cM}lV_{fM}\right.\nonumber\\
&\left.-\frac{1}{2}V_{iM}l^2V_{fM}V_{iM}lV_{fM}-\frac{1}{2}V_{iM}l^2V_{cM}lV_{fM}V_{0M}\right. \nonumber\\
&+\left.-\frac{1}{2}V_{iM}lV_{cM}l^2V_{fM}V_{0M}\frac{1}{2}V_{iM}l^3V_{fM}V_{0M}^2 + h.c.\right) + \cdots\nonumber\\.
\end{align}
Keeping the lowest two orders, the effective Hamiltonian becomes
\begin{equation}
H_{\text{eff}} = J \left[1+\frac{\left(\Omega-\Omega'\right)\left(\Omega^{'*}-\Omega^*\right)}{\delta\omega^2}\right] \sigma_x,
\end{equation}
where $\sigma_x$ is the Pauli matrix. Therefore, the bare DDI is corrected by the Raman potential through $J_{\text{DDI}} = J [1+2\delta\omega^{-2}\left(\Omega-\Omega'\right)(\Omega^{'*}-\Omega^*)]$. Including correction to all orders, the bare DDI is then corrected by $J_\text{DDI}^{\left(\infty\right)} = JJ_0\left(2\delta\omega^{-2}\left(\Omega-\Omega'\right)(\Omega^{'*}-\Omega^*)\right)$, where $J_0(\cdot)$ is the Bessel function of the first kind.

The Floquet Hamiltonian for LADDI can be similarly written as (there is detuning $\Delta$ at $\vec{r}_2$ and no detuning at $\vec{r}_1$)
\begin{equation}
H_{T,\text{LADDI}} = \left(
\begin{array}{ccccc}
\ddots & \ddots & \ddots & &\\
\ddots & \mathcal{D}-\delta\omega & \mathcal{Q}' & 0 &\\
\ddots & \mathcal{Q}'^\dagger & \mathcal{D} & \mathcal{Q}' & \ddots\\
& 0 & \mathcal{Q}'^\dagger & \mathcal{D}+\delta\omega & \ddots\\
& & \ddots & \ddots & \ddots\\
\end{array}
\right),
\end{equation}
where $\mathcal{D} = \big(\begin{smallmatrix} \delta & 0\\0 & 0  \end{smallmatrix}\big)$, $\mathcal{Q}' =  \big(\begin{smallmatrix} \Omega' & 0\\J & \Omega   \end{smallmatrix}\big)$, $\delta$ is an extra detuning to compensate the $AC$ stark shift due to the non-resonant bare DDI. To the lowest order, $\delta = -2J^2/\delta\omega$, and the effective Hamiltonian to the lowest order is
\begin{equation}
H_{\text{eff, LADDI}}  = \left(
\begin{array}{cc}
0 & \frac{J}{\delta\omega}\left(\Omega-\Omega'\right)\\
\frac{J}{\delta\omega}\left(\Omega^*-\Omega^{'*}\right) & 0
\end{array}
\right).
\end{equation}
Therefore, to the lowest order, $J_\text{LADDI} = J\delta\omega^{-1} \left(\Omega(\vec{r}_1) - \Omega(\vec{r}_2)\right)$. The effective LADDI, including infinite order corrections, can be similarly obtained $J^{\left(\infty\right)}_\text{LADDI} = J\frac{\Omega(\vec{r}_1) - \Omega(\vec{r}_2)}{\left| \Omega(\vec{r}_1) - \Omega(\vec{r}_2)\right|} J_1\left(2\delta\omega^{-1}\left|\Omega(\vec{r}_1) - \Omega(\vec{r}_2)\right|\right)$.

\section*{Appendix C: Experimental parameters}
In this section, we give the estimate of the relevant experimental parameters. To be specific, the data given here are based on $^{87}$Rb atoms. Choosing $a_x$ to be $14\mu$m, the typical size of DDI would be in the order of $J_0 = \left|C_3 a_x^{-3}\right| = 1.1$ MHz. To observe the correlated effects, the lifetime of the system should be large compared to the characteristic time of the system. For $J_0 = 1.1$MHz, a lifetime of at least $\tau> \tau_{0} = 50J_0^{-1} = 60\mu s$ is desirable. Spontaneous emissions lead to decay of the Rydberg states $\left|0\right>$ and $\left|1\right>$ and the $6P$ states, which the Raman process coupled. The estimated lifetimes of $\left|0\right>$ and $\left|1\right>$ are $500\mu$s and $230\mu$s respectively, which are much longer than the required lifetime $\tau_0$; the lifetime of $6P$ states has a lifetime of $\tau_{6P}=130 ns$. So for the single photon detuning $\Delta_R = 18.5$ GHz and two photon detuning $\Delta = 6.6$ MHz, the corresponding Rabi frequency should satisfy $\left(2|\tilde{\Omega}|/\Delta_R\right)^{-2} \tau_{6P} > \tau_0$, i.e. $|\tilde{\Omega}|<500$ MHz. For the required Raman potential $\Omega_1 \approx \Omega_2 \approx 3.3$MHz, the corresponding Rabi frequencies $|\tilde{\Omega}_1| \approx |\tilde{\Omega}_2| \approx 250$MHz satisfy the requirement for $\tau$.

\section*{Appendix D: Many body Chern number}
In this section, we introduce the details of obtaining many-body Chern
number~\cite{MBCN1,MBCN2}. For a many-body state $|\Psi\rangle$, the twisted boundary
conditions are defined such that
\begin{align*}
 & \langle\boldsymbol{r}+N_{x}\boldsymbol{x}|\Psi\rangle=e^{i\vartheta_{x}}\langle\boldsymbol{r}|\Psi\rangle,\\
 & \langle\boldsymbol{r}+N_{y}\boldsymbol{y}|\Psi\rangle=e^{i\vartheta_{y}}\langle\boldsymbol{r}|\Psi\rangle.
\end{align*}
Recall that for the non-degenerate case, the Chern number $C$ for $\alpha$th many-body eigenstate of Hamiltonian is given by
\begin{equation}
  C(\alpha)=\frac{1}{2\pi}\int_{0}^{2\pi}d \vartheta_x  \int_{0}^{2\pi}d\vartheta_y (\partial_x \mathcal{A}_{y}^{\alpha }-\partial_{y}\mathcal{A}_{x}^{\alpha}),
\end{equation}
where $\mathcal{A}_{j}^{\alpha}(\vartheta _{x},\vartheta _{y})=i \langle \Psi ^{(\alpha)} \rvert \partial_{\vartheta _{j}}\lvert \Psi ^{(\alpha)} \rangle$ is defined as a vector field based on the eigenstate $\Psi^{(\alpha)}(\vartheta _{x},\vartheta _{y})$.
For the degenerate ground states, we define instead the tensor field
\begin{equation}
\mathcal{A}_{j}^{(\alpha,\beta)}(\vartheta_{x},\vartheta_{y})=i\langle\Psi^{(\alpha)}|\frac{\partial}{\partial\vartheta_{j}}|\Psi^{(\beta)}\rangle,
\end{equation}
where $\alpha,\beta=1,2$ are indices for two degenerate states. And the many-body Chern number shared by ground states multiplets is
\begin{equation}
   C(\alpha,\beta)=\frac{1}{2\pi}\int_{0}^{2\pi}d \vartheta_x  \int_{0}^{2\pi}d\vartheta_y \mathrm{Tr}(\partial_x \mathcal{A}_{y}^{\alpha,\beta }-\partial_{y}\mathcal{A}_{x}^{\alpha,\beta}),
\end{equation}
which is equivalent to the path integral around the singularity of tensor field $\mathcal{A}_{j}^{\alpha,\beta }(\vartheta_{x},\vartheta _{y})$. From a practical point of view, we choose the two ground states multiplets with twist angles far from each other, for example $\lvert\Phi^{(\alpha)}\rangle = \lvert\Psi^{(\alpha)}(\vartheta_x=0,\vartheta_y=0)\rangle$, $\lvert\Phi'^{(\alpha)}\rangle=\lvert\Psi^{(\alpha)}(\pi,\pi)\rangle$ as gauge references, and define the scalar overlapping field
\begin{equation}
\Delta_\Phi(\vartheta_{x},\vartheta_{y})=\det\left(\begin{matrix}
\langle\Phi^{(1)}|P|\Phi^{(1)}\rangle & \langle\Phi^{(1)}|P|\Phi^{(2)}\rangle \\
\langle\Phi^{(2)}|P|\Phi^{(1)}\rangle &
\langle\Phi^{(2)}|P|\Phi^{(2)}\rangle
\end{matrix}\right),
\end{equation}
where $P=|\Psi^{(1)}(\vartheta_{x},\vartheta_{y})\rangle\langle\Psi^{(1)}(\vartheta_{x},\vartheta_{y})|+|\Psi^{(2)}(\vartheta_{x},\vartheta_{y})\rangle\langle\Psi^{(2)}(\vartheta_{x},\vartheta_{y})|$ is the projection operator on the ground state manifold. Then the Chern number for two degenerate states will equal to the number of vortices of
$\Omega(\vartheta_{x},\vartheta_{y})$ field where $\Delta_\Phi$ (or
$\Delta_{\Phi'}$) vanishes (i.e., singularities of the tensor field)~\cite{MBCN1,MBCN2}, with $\Omega(\vartheta_{x},\vartheta_{y})$ defined as
\begin{equation}
\Omega(\vartheta_{x},\vartheta_{y})=\det\left(\begin{matrix}
\langle\Phi'^{(1)}|P|\Phi^{(1)}\rangle & \langle\Phi'^{(1)}|P|\Phi^{(2)}\rangle \\
\langle\Phi'^{(2)}|P|\Phi^{(1)}\rangle &
\langle\Phi'^{(2)}|P|\Phi^{(2)}\rangle
\end{matrix}\right).
\end{equation}
As shown in Figure~\ref{fig:ED Fig}(d) in main text, such vorticity equals to 1, indicating the each ground state has a Chern
number $C=\frac{1}{2}$, as expected for the $\nu=\frac{1}{2}$ fractional quantum anomalous Hall phase.

\begin{figure*}[htp]
\centering\includegraphics{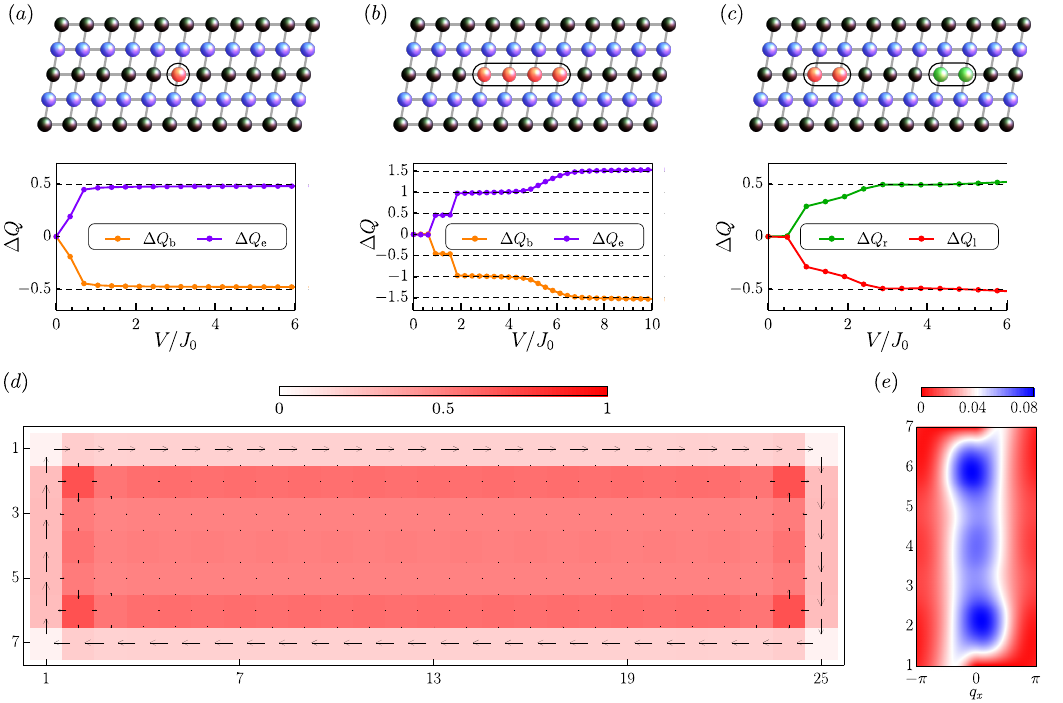}\caption{\label{fig:SM}Net charge $\Delta Q_{e(b)}$ of the edge (bulk) as a function of $V$ on a $L_x=25$, $L_y=5$ superarray and $\nu=1/2$, with single (a) or four (b) repulsive potential $V$ (red sphere) introduced at the center of the superarray. (c) $\Delta Q$ for the system in the presence of both repulsive (red spheres) and attractive potentials (green spheres). The repulsive and attractive potentials are introduced at the left and right half center of the system, respectively. As the $V$ increases, the net charge saturates to different values with quantized step $1/2$.  (d) Density and local current profiles for $L_x=25$, $L_y=7$ and $\nu=1/2$. The color represents local density, length and arrows represent magnitudes and directions of local currents, respectively. Chiral edge current is observed. (e) Momentum distribution $n_{j_y}(q_{x})$. $n_{j_y}(q_{x})$ has a finite occupation around $q_{x}=q_{0}(-q_{0})\protect\neq0$
on $j_y=1$ ($j_y=7$) while it is peaked around $q_{x}=0$ in the bulk chains, indicating the existence of opposite chiral current in opposite edges.}
\end{figure*}

\section*{Appendix E: Charge fractionalization}
In this section, we provide more numerical results for charge fractionalization as complement to the main text. We compute the net charge as a function of potential strength $V$ on a $25\times5$ Rydberg superarray with open boundary condition by density-matrix renormalization group (DMRG) simulations with bond dimension up to $\chi\approx600$. We consider the case of one site defect and four sites defect here. As shown in \textbf{Figure~\ref{fig:SM}}(a)-(b), the net charges $\Delta Q_{\mathrm{e}(\mathrm{b})}$ saturate different values for different defect configuration. For a superarray with size $L_x=25$ and $L_y=5$, $\Delta Q_{\mathrm{e}(\mathrm{b})}$ saturates to $\pm 1/2$ if the defect occupies one site, while saturates to $\pm 3/2$ with quantized step $1/2$ if the defect occupies four sites.

We also consider the case that system in the presence of both repulsive and attractive defects. The repulsive potential
is placed on $(6,3)$ and $(7,3)$ and the attractive potential is
placed on $(18,3)$ and $(19,3)$, respectively. Then the net charge of left half $\Delta Q_{\mathrm{l}}$ and right half $\Delta Q_{\mathrm{r}}$ of the system are shown in Figure~\ref{fig:SM}(c). After the strength of pin filed is above a threshold, the net
charge saturates to quantized values close to $\pm1/2$ as expected.

\section*{Appendix F: More Experimental observables}
In this section, we examine more observables accessible to experiments.
More specifically, we present local particle density, the chiral edge
currents and momentum distributions of our system. Throughout this
section, the calculation is performed on open boundary condition by DMRG simulations. We first examine the local particle density defined as
\begin{equation}
\begin{cases}
n_{j_{x},j_{y}}=\langle a_{j_{x},j_{y}}^{\dagger}a_{j_{x},j_{y}}\rangle, & j_{x}\text{ is odd},\\
n_{j_{x},j_{y}}=\langle b_{j_{x},j_{y}}^{\dagger}b_{j_{x},j_{y}}\rangle, & j_{x}\text{ is even},
\end{cases}
\end{equation}
and the currents from different hopping processes defined as
\begin{align}
I_{j_{x},j_{y}}^{x} & =-iJ_{x}\langle a_{j_{x},j_{y}}^{\dagger}b_{j_{x}+1,j_{y}}\rangle+c.c.,\\
I_{j_{x},j_{y}}^{\mathrm{nn}x} & =-iJ_{\mathrm{nnx}}\langle a_{j_{x},j_{y}}^{\dagger}a_{j_{x}+2,j_{y}}\rangle+c.c.,\\
I_{j_{x},j_{y}}^{y} & =-iJ_{y}e^{i\pi j_{x}}\langle a_{j_{x},j_{y}}^{\dagger}a_{j_{x},j_{y}+1}\rangle+c.c.,
\end{align}
\begin{align}
I_{j_{x},j_{y}}^{\mathrm{nn}y} & =-iJ_{\mathrm{nny}}e^{i\pi j_{x}}\langle a_{j_{x},j_{y}}^{\dagger}a_{j_{x},j_{y}+2}\rangle+c.c.,\\
I_{j_{x},j_{y}}^{d_{1}} & =-|J_{d}|e^{i\pi j_{x}}\langle a_{j_{x},j_{y}}^{\dagger}b_{j_{x}+1,j_{y}+1}\rangle+c.c.,\\
I_{j_{x},j_{y}}^{d_{2}} & =-|J_{d}|e^{i\pi j_{x}}\langle b_{j_{x}+1,j_{y}}^{\dagger}a_{j_{x},j_{y}+1}\rangle+c.c.,
\end{align}
for odd $j_{x}$, and $a_{j_x,j_y}\leftrightarrow b_{j_x,j_y}$ for even $j_{x}$. The
density profiles and local currents are shown in Figure~\ref{fig:SM}(d).
The color indicates the magnitude of $n_{j_{x},j_{y}}$. The length
and the arrows represent the strength and the direction of current
along a bond, respectively. The density profile shows a spike pattern
at the second ring, resembling the density profile of a bosonic $1/2$-Laughlin state~\cite{Haldane,Jaksch}.
The current shows a clear peaking at the edge and decay quickly into
bulk, with opposite directions on the different edges~\cite{Grusdt},
manifesting a chiral edge currents and insulating bulk states in the
ground state.

We further examine the (quasi-)momentum distribution along $x$-direction
in a given $j_{y}$,
\begin{equation}
n_{j_{y}}(q_{x})=\frac{1}{L_{x}}\sum_{j_{x},j_{x}'}e^{-iq_{x}(j_{x}-j_{x}')}\langle d_{j_{x},j_{y}}^{\dagger}d_{j_{x}',j_{y}}\rangle,
\end{equation}
with $d_{j_{x},j_{y}}=a_{j_{x},j_{y}}$ ($b_{j_{x},j_{y}}$) for odd
(even) $j_{x}$, and $q_{x}=2\pi n/L_{x}$, with $n=0,1,...,L_{x}-1$.
The (quasi-)momentum profiles (Figure~\ref{fig:SM}) show
a finite occupation around $q_{x}=q_{0}(-q_{0})\neq0$ on the edge
and peaks around $q_{0}=0$ in the bulk, consistent with chiral nature
of the edge current.


\begin{thebibliography}{100}
\bibitem{detuneNote}
It is assumed that such detuning $\Delta$ is realized by applying AC stark
  shift on the Rydberg states $\left|1B\right>$ for more simple discussion, but
  we emphasis here that such method is not the one and only way to realize such
  detuning. For example, if we choose the Rydberg states $\left|0(1),A\right> =
  \left|nS(P)\right>$ and $\left|0(1),B\right> = \left|n'S(P)\right>$ with
  $\left|n-n'\right|\geq 3$, the detuning $\Delta$ is naturally realized since
  the energy differences between $S$ and $P$ states with different principal
  quantum numbers $n$ are different. The choice $\left|n-n'\right|\geq 3$ is
  necessary to suppressed the unwanted bare DDI between $\left|0A\right>$ and
  $\left|1B\right>$ or $\left|1A\right>$ and $\left|0B\right>$. Another way is
  to use different atoms for different sublattices so that the detuning is
  realized through different energy spacings for different atoms.
\end{thebibliography}
\end{document}